\documentclass[conference]{IEEEtran}
\IEEEoverridecommandlockouts
\usepackage{hyperref}
\hypersetup{colorlinks,allcolors=black}
\usepackage{cite}
\usepackage{amsmath,amssymb,amsfonts}
\usepackage{algorithmic}
\usepackage{graphicx}
\usepackage{textcomp}
\usepackage{xcolor}
\usepackage{float}
\usepackage{diagbox}
\usepackage{soul}
\usepackage{makecell}
\def\BibTeX{{\rm B\kern-.05em{\sc i\kern-.025em b}\kern-.08em
    T\kern-.1667em\lower.7ex\hbox{E}\kern-.125emX}}

\begin{document}

\title{Benchmarking of Lightweight Deep Learning Architectures for Skin Cancer Classification using ISIC 2017 Dataset}

\makeatletter
\newcommand{\linebreakand}{%
  \end{@IEEEauthorhalign}
  \hfill\mbox{}\par
  \mbox{}\hfill\begin{@IEEEauthorhalign}
}

\author{\IEEEauthorblockN{1\textsuperscript{st} Abdurrahim Yilmaz}
\IEEEauthorblockA{\textit{Yildiz Technical University} \\
\textit{Mechatronics Engineering}\\
Istanbul, Turkey \\
a.rahim.yilmaz@gmail.com}
\and
\IEEEauthorblockN{2\textsuperscript{nd} Mucahit Kalebasi}
\IEEEauthorblockA{\textit{Yildiz Technical University} \\
\textit{Mechatronics Engineering}\\
Istanbul, Turkey \\
mucahitkalebasi@gmail.com}
\and
\IEEEauthorblockN{3\textsuperscript{rd} Yegor Samoylenko}
\IEEEauthorblockA{\textit{Yildiz Technical University} \\
\textit{Mechatronics Engineering}\\
Istanbul, Turkey \\
yegoraleksandr@gmail.com}
\linebreakand
\IEEEauthorblockN{4\textsuperscript{rd} Mehmet Erhan Guvenilir}
\IEEEauthorblockA{\textit{Hisar School} \\
\textit{}\\
Istanbul, Turkey \\
mehmet.guvenilir@hisarschool.k12.tr}
\and
\IEEEauthorblockN{5\textsuperscript{th} Huseyin Uvet}
\IEEEauthorblockA{\textit{Yildiz Tehnical University} \\
\textit{Mechatronics Engineering}\\
Istanbul, Turkey \\
huvet@yildiz.edu.tr}
}

\maketitle

\begin{abstract}
Skin cancer is one of the deadly types of cancer and is common in the world. Recently, there has been a huge jump in the rate of people getting skin cancer. For this reason, the number of studies on skin cancer classification with deep learning are increasing day by day. For the growth of work in this area, the International Skin Imaging Collaboration (ISIC) organization was established and they created an open dataset archive. In this study, images were taken from ISIC 2017 Challenge. The skin cancer images taken were preprocessed and data augmented. Later, these images were trained with transfer learning and fine-tuning approach and deep learning models were created in this way. 3 different mobile deep learning models and 3 different batch size values were determined for each, and a total of 9 models were created. Among these models, the NASNetMobile model with 16 batch size got the best result. The accuracy value of this model is 82.00\%, the precision value is 81.77\% and the $F_1$ score value is 0.8038. Our method is to benchmark mobile deep learning models which have few parameters and compare the results of the models. 
\end{abstract} 


\section{Introduction}
Skin cancer is a dangerous and a common type of cancer that exist in our today's world. Sun is important for human skin and body but comes with ultraviolet (UV) radiation which is the main source of skin cancer. Due to UV radiation, the melanocyte, a type of skin cell, begin to develop uncontrollably and becomes skin cancer lesion. One in three newly diagnosed malignancies is a skin cancer \cite{laikova2019advances}. In the last decade the amount of skin cancer cases increased by nearly 50\%. The high percentage of increase shows the importance of working in the field of skin cancer and its early detection \cite{oliveira2018computational}. 
Melanoma is the term used to describe a type of skin cancer categorized as a malignant tumor caused by melanocytes. Melanoma is responsible for more than 75\% of skin cancer-related deaths. When skin cancer lesions appear, certain patterns also occur. These patterns can be seen better with the dermoscopes, but a dermatologist is still needed for diagnosis. However, the procedure of diagnosing skin cancer can become lengthy in economically under-developed and developing regions that lack well-trained dermatologists. To contribute to this circumstance, research on deep learning models capable of detecting skin cancer has begun.
Dermatology is a field of medicine that focuses on skin disorders, and dermatologists are experts in the field. Dermatologists primarily examine the lesion with the naked eye. However, for further examination, they require dermoscopy devices called dermoscopes, which contain a lens and an polarized or non-polarized light source for magnification. These devices make superficial and deeper patterns of lesions noticeable and enables data collection. Dermoscopy devices come in fixed and hand-held variants. Fixed devices also can be called digital dermoscopy devices, such as Molemax, consist of a computer, camera and a dermoscope. Handhelds are small, designed to be portable, consist of dermoscope and may not come with camera. Due to the design, the mobile phones can also be used as a image capturing devices.
Literature provides lots of research and studies of different deep learning approaches on skin cancer and ISIC one of them. The ISIC archive serves as a primary dataset for studies on skin cancer. The ISIC archive contains skin cancer datasets from more than one institution and grows by incorporating different datasets every year. The annual competitions organized by ISIC consist of three parts: lesion segmentation, dermoscopic feature classification, and disease classification. Participants are asked to create a deep learning model that automatically predicts lesion segmentation from dermoscopic images with binary masks in the lesion segmentation part. In the dermoscopic feature classification section, participants are asked to identify four clinically defined dermoscopic features automatically. In the disease classification section, participants are asked to classify images according to cancer types \cite{berseth2017isic}. 
Mohammed et al. \cite{al2018skin} gather the dataset consisting of 2750 images from the ISIC 2017 Challenge. The training dataset contains 2000 images, the validation dataset contains 150 images, and the test dataset contains 600 images.  For data augmentation, they used the Hue-Saturation-Value (HSV) format and image rotations. They added the images in the dataset in HSV format and reached 4000 images in total. The images in the dataset were rotated 0$^{\circ}$, 90$^{\circ}$, 180$^{\circ}$ and 270$^{\circ}$ to increase the dataset to 16000 images. As a result, 16000 dermoscopy images were used while training the proposed full resolution convolutional networks (FrCN) segmentation method. The pre-trained VGG16 model was used in model training. Their FrCN outperformed the others with overall accuracy, dice, and Jaccard indices of 94.03\%, 87.08\%, and 77.11\%, respectively. In contrast, U-Net achieved the highest specificity for overall skin lesion segmentation at 97.24\%. The segmentation accuracies of the benign, melanoma and SK cases in the ISIC 2017 test dataset for their proposed FrCN were 95.62\%, 90.78\%, and 91.29\%, respectively. The computational speed during the training phase of their proposed FrCN method was faster than other segmentation methods.
Jeremy et al. \cite{kawahara2018fully} created dermoscopic images using the ISIC 2017 Challenge image archive. They recast the classification of clinical dermoscopic features contained within superpixels as a segmentation problem and propose a fully convolutional neural network for detecting clinical dermoscopic features included within dermoscopy skin lesion images. They expand VGG16, a convolutional neural network pre-trained on ImageNet, by implementing a semantic segmentation architecture comparable to VGG16. They remove fully connected layers of VGG16 and use bilinear interpolation to resize chosen responses and feature maps throughout the network to match the size of the input image. Concatenation of these enlarged feature maps enables them to evaluate feature maps from several network levels directly. When compared to the other models, their technique had the highest mean AUC score. Especially, they achieve the highest results across all measures for the pigment network dermoscopic characteristic. Additionally, they experiment with replacing VGG16 with more contemporary models, such as ResNet50 and InceptionResNetV2. They discovered that altering the underlying model did not affect the outcome.
Philipp et al. \cite{tschandl2019diagnostic} gathered 888, 2750, and 16691 photos, respectively, from the EDRA, ISIC2017, and PRIV datasets. They aimed to evaluate the diagnostic accuracy of the Content-Based Image Retrieval (CBIR) method to that of neural network predictions. In all three datasets, CBIR forecasts of skin cancer had AUC values similar to those of neural networks. When employed on an eight-class dataset, neural networks trained to detect only three classes outperformed CBIR. On the EDRA dataset, the neural network performed the best, predicting with a 76.2\% accuracy.
The significance of this article is to benchmark mobile deep learning models with different batch size parameters on ISIC 2017 dataset and show how they perform. We used MobileNet, MobileNetV2 and NASNetMobile architectures to perform mobile deep learning models. The dataset is resized to 224$\times$224$\times$3 and is saved as numpy arrays for time efficiency. Data augmentation is applied to overcome overfitting. The best accuracy and precision values that were acquired for ISIC 2017 dataset classification NASNetMobile model with 16 batch size was with accuracy of 82\%.
\section{Methodology}
Artificial intelligence (AI) is a broad field ranging from fundamental regression of numbers to autonomous driving and decision making. Deep learning is a type of AI that learns features after being fed data, thus imitating humans. Usage of deep learning in the medical sphere mainly comes as classification and segmentation. These methods use Convolutional Neural Networks (CNN) as a base for working with digital images. CNN with each convolution layer learns features. With the increasing amount of convolution layers, layers begin to learn higher-level features. Transfer learning is an approach that takes previously trained models and uses them on new datasets \cite{olivas2009handbook}. With transfer learning, learning starts from a higher accuracy due to previously learned features. Features can be low or high level based on the length of transferred model. While common features can be used nearly on every data, high-level features should be used only on a similar dataset on which the transferred model was trained. ImageNet is generally used for training models. The huge variety and a large amount of data enable models to learn a good portion of features. Imagenet is a large dataset consisting of 21,841 classes and 14,197,122 images \cite{deng2009imagenet}. 
Typical usage of transfer learning consists of loading a model trained on the ImageNet dataset, making layers non-trainable, removing the last dense layer, adding a new dense layer, and then initializing the training. The transfer learning method predicts the shapes, colors, and texture structure learned by a pre-trained deep learning model on the images in the new dataset. The last layer of the frozen pre-trained model is removed and re-added accordingly to the new dataset. However, model accuracy may not improve after a while if the previously mentioned dataset issue exists. The fine-tuning method is based on retraining the pre-trained model by opening the determined layers to training and updating the layers according to the content of the new dataset. To further improve accuracy, fine-tuning named training method is used. This method is applied by making the following changes; After training, previously made non-trainable layers are made trainable, and the learning rate is lowered. With fine-tuning, features will be optimized for the current dataset, and accuracy may rise significantly. Of course, the increase is not guaranteed as accuracy depends on the model used, the dataset used, and the hyperparameters used. Our study uses MobileNet, MobileNetV2, and NASNetMobile with transfer learning. All models were pre-trained on the ImageNet dataset and were trained on the ISIC2017 dataset.

\subsection{Dataset}
ISIC 2017 dataset consists of 2750 skin cancer images. The images consist of 2000 training datasets, 150 test datasets, and 600 validation datasets. The images range in size from 540$\times$722$\times$3 to 4499$\times$6748$\times$3. High resolution of images in the ISIC dataset provides an advantage over other dermoscopic image datasets. Although the number of images in the dataset is not high enough and the dataset is weak in terms of species diversity, the deep learning model achieved high success in experimental settings, although it achieved moderate success in the clinical environment \cite{berseth2017isic}.

\subsection{Models}
\subsubsection{MobileNet}
MobileNet is a deep learning model that was designed to be light and have low latency for mobile and embedded devices. MobileNet has 88 layers, 4.2 million parameters and has a size of only 16 MB, its size is very small compared to other models. MobileNet architecture is built on Depthwise Seperable Convolutions. Depthwise Seperable Convolutions consists of two layers, one being depthwise convolution and the other one being pointwise convolution. Depthwise applies single filter per input depth, pointwise is used to create linear combination of output from the depthwise layer. Depthwise convolution is more efficient, uses 8 to 9 times less computation, compared to normal convolution layers \cite{howard2017mobilenets}.

\subsubsection{MobileNetV2}
This deep learning model, inherits the goals of MobileNet and further improves on them. MobileNetV2 has a lower amount of layers than MobileNet and uses even less computational power. Amount of layers is same as MobileNet which is 88 layers, amount of parameters decreases to 3.5 million parameters and size also decreases to 14 MB. With MobileNetV2, they improve MobileNet by adding, shortcut connections, inverted residual blocks and bottleneck blocks. Inverted residual bottleneck layers allows to have a memory efficient implementation. This model also does a great job in object detection and semantic segmentation \cite{sandler2018mobilenetv2}.

\subsubsection{NASNetMobile}
This model, was created by Google via NASNet search space, named the way it is due to giving rise to NASNet, with the goal of finding the best parameters and creating the best model. NAS stands for Neural Architecture Search, automates the search and finds the best algorithm. Although the results are impressive applying NAS to a large dataset is not efficient in terms of computation. While NASNetLarge has nearly 89 million parameters NASNetMobile only has 5.3 million parameters. Same is said for the model sizes as the NASNetLarge is 343 MB, NASNetMobile is only 23 MB \cite{zoph2018learning}.

\subsection{Implementation}
In the deep learning method, the high-resolution images filling the computer memory is one of the main reasons that affect the duration of the training. Therefore, images from the ISIC archive are scaled to 224$\times$224. The photos with reduced resolution were saved by converting the NumPy library to array format. Following that, training was conducted using these NumPy arrays. MobileNet, MobileNetV2, and NASNetMobile architectures are trained with three different batch size hyperparameters for benchmark testing. The GlobalAveragePooling layer has been added to the end of the architectures. The flatten layer converts the image derived from the architecture into a format suitable for processing in the fully connected layer. In addition, three different dense layers were added with a dropout value of 0.2. The output layer containing the number of classes has been added at the end of the architecture. Training and testing were performed on a system with following components; Ryzen ThreadRipper 1950X, Nvidia GTX 1080Ti,32 GB of RAM and 512 GB SSD. While time required for one epoch is relatively small, like 20 seconds averagely, due to high epoch count the overall time required was 8 hours to complete.

Pre-trained models on the ImageNet dataset were used with a 2-step training for fine-tuning the model. In the first step, the layers of the models are frozen. Only the fully connected block added to the end of the architecture is trained. Thus, the selection was made according to the universal features extracted from the ImageNet dataset. In the second stage, fine-tuning was done. At this stage, all model layers were included in the training. The universal features extracted from the ImageNet dataset were optimized to obtain the model with the highest performance. Model training was realized using the Keras library.

\begin{table}[h!]
\renewcommand{\arraystretch}{1.25}
\centering
\caption{General Properties of the Models}
\begin{tabular}{ccc}
\hline\hline      
\textbf{Property}      & \textbf{Value}      \\ 
\hline  
Dataset Size           & 2750                           \\
Image Shape           & (224,224,3)                     \\
Test Data/All Data   & 0.15                             \\
Dropout                     & 0.2                       \\
2D Pooling Size             & 2$\times$2                \\
Validation Data/Training Data  & 0.20                   \\
Epoch Number         & 150                              \\
\hline\hline
\end{tabular} \label{tab:table1}
\end{table}

\subsection{Metrics}
Metrics are used to evaluate and compare the performance of models. The metrics show accuracy, sensitivity, specificity, $F_1$ score and area under the curve (AUC) score. These parameters can be calculated by used confusion matrix. The formulas for these classification metrics are shown sequentially in Table \ref{tab:table1}. The receive operating characteristics (ROC) curve is another test tool for determining how probabilistically the model provides results. It is a curve that shows the ratio of true to false predictions when the ROC curve is used to calculate the threshold values for each class and the decision is made using the determined threshold value for that class.

\begin{table}[h!]
\renewcommand{\arraystretch}{1.25}
\centering
\caption{Metrics}
\begin{tabular}{ccc} 
\hline\hline
\textbf{Metric}  & \textbf{Formula}              \\ 
\hline \\ [-10pt]
Accuracy       &   $\dfrac{T P+T N}{T P+F P+T N+F N}$  \\ [5pt]
Sensitivity  & $\dfrac{T P}{T P+F N}$ \\ [5pt]
Precision  & $\dfrac{T P}{T P+F P}$ \\ [5pt]
Specificity  & $\dfrac{T N}{T N+F P}$ \\ [5pt]
$F_1$ Score  & $2*\dfrac{\text{Precision} \times \text{Sensitivity}}{\text{Precision}+\text{Sensitivity}}$ \\ [5pt]
AUC  & Area under of the curve \\ [5pt]
\hline\hline
\end{tabular} \label{tab:table2}
\end{table}

\section{Results and Discussion}
In this study, a performance comparison of deep learning models for skin cancer classification was made. Table \ref{tab:table3} shows the performance of the trained deep learning models on the ISIC 2017 dataset. MobileNet model with 16 batch size has 80.36\% accuracy, 79.38\% precision and $F_1$ score with 0.7934 results. The MobileNet model with 32 batch size has 80.73\% accuracy, 80.72\% precision and 0.8057 $F_1$ score. The MobileNet model with 64 batch size has 77.64\% accuracy, 76.26\% precision and 0.7567 $F_1$ score. MobileNetV2 model with 16 batch size has 81.45\% accuracy, 81.28\% precision and 0.8114 $F_1$ score. The MobileNetV2 model with 32 batch size has 78.55\% accuracy, 78.85\% precision and 0.7806 $F_1$ score. The MobileNetV2 model with 64 batch size has 80.91\% accuracy, 80.19\% precision and 0.8035 $F_1$ score. NASNetMobile model with 16 batch size has 82\% accuracy, 81.77\% precision and 0.8038 $F_1$ score. The NASNetMobile model with 32 batch size has 79.64\% accuracy, 79.38\% precision and 0.7823 $F_1$ score. The NASNetMobile model with 64 batch size has 77.45\% accuracy, 77.94\% precision and 0.7510 $F_1$ score.

The NASNetMobile model with 16 batch size got the best accuracy and precision values when benchmarking between pre-trained models. This accuracy, precision, and $F_1$ score values of the model are 82.00\%, 81.77\%, 0.8038, respectively. According to the model results, models with a small batch size performed significantly better in class generalization. The average accuracy of the MobileNet, MobileNetV2, and NASNetMobile models is 79.60\%, 80.30\%, and 79.70\%, respectively. Looking at the average accuracy values, we see that the MobileNetV2 models have better performance.

\begin{table}[h!]
\renewcommand{\arraystretch}{1.25}
\centering
\caption{Results}
\begin{tabular}{c|c|c|c|c}
\hline\hline      
\textbf{Model - Batch Size}  & \textbf{Accuracy} & \textbf{Precision} &  \textbf{$F_1$ Score} &  
\textbf{\makecell{AUC\\ Score}}\\ 
\hline  
MobileNet - 16               &  80.36\% &        79.38\% &          0.7934      &  \textbf{0.9050}                       \\
MobileNet - 32               &  80.73\% &        80.72\% &          0.8057      &  0.8936                      \\
MobileNet - 64               &  77.64\% &        76.26\% &          0.7567      &  0.8883                \\
MobileNetV2 - 16             &  81.45\% &        81.28\% &  \textbf{0.8114}     &  0.9035                    \\
MobileNetV2 - 32             &  78.55\% &        78.85\% &          0.7806      &  0.8977                       \\
MobileNetV2 - 64             &  80.91\% &        80.19\% &          0.8035      &  0.9023                  \\
\textbf{NASNetMobile - 16}   &  \textbf{82.00\%} &  \textbf{81.77\%} &  0.8038  &  0.8832                     \\
NASNetMobile - 32            &  79.64\% &        79.38\% &          0.7823      &  0.8816                   \\
NASNetMobile - 64            &  77.45\% &        77.94\% &          0.7510      &  0.8684                   \\
\hline\hline
\end{tabular} \label{tab:table3}
\end{table}
\section{Conclusion}

In this study, three different deep learning models were used with transfer learning and fine-tuning approach. Three different batch size values were used and in total nine models were created. Skin cancer disease classification was made with mobile deep learning models determined using ISIC 2017 dataset, and the performances of these models were benchmarked. Since mobile deep learning models have fewer parameters than other models, the training and prediction times were shorter. Although a deep learning model generally gives a good result on a specific dataset, hyperparameter optimization is required to find the best result. In this study, it is compared to what extent batch size values, one of the hyperparameters in deep learning models, will affect the generalization ability of these models. Models with a batch size of 16 gave good results, while models with a batch size of 64 gave poor results in general.

\bibliographystyle{ieeetr}
\bibliography{reference.bib}

\cleardoublepage
\pagebreak
\onecolumn
\section*{SUPPLEMENTARY}

\begin{figure}[h!]
\centering
\label{fig:mobilenetv1_16}
\includegraphics[width=8cm]{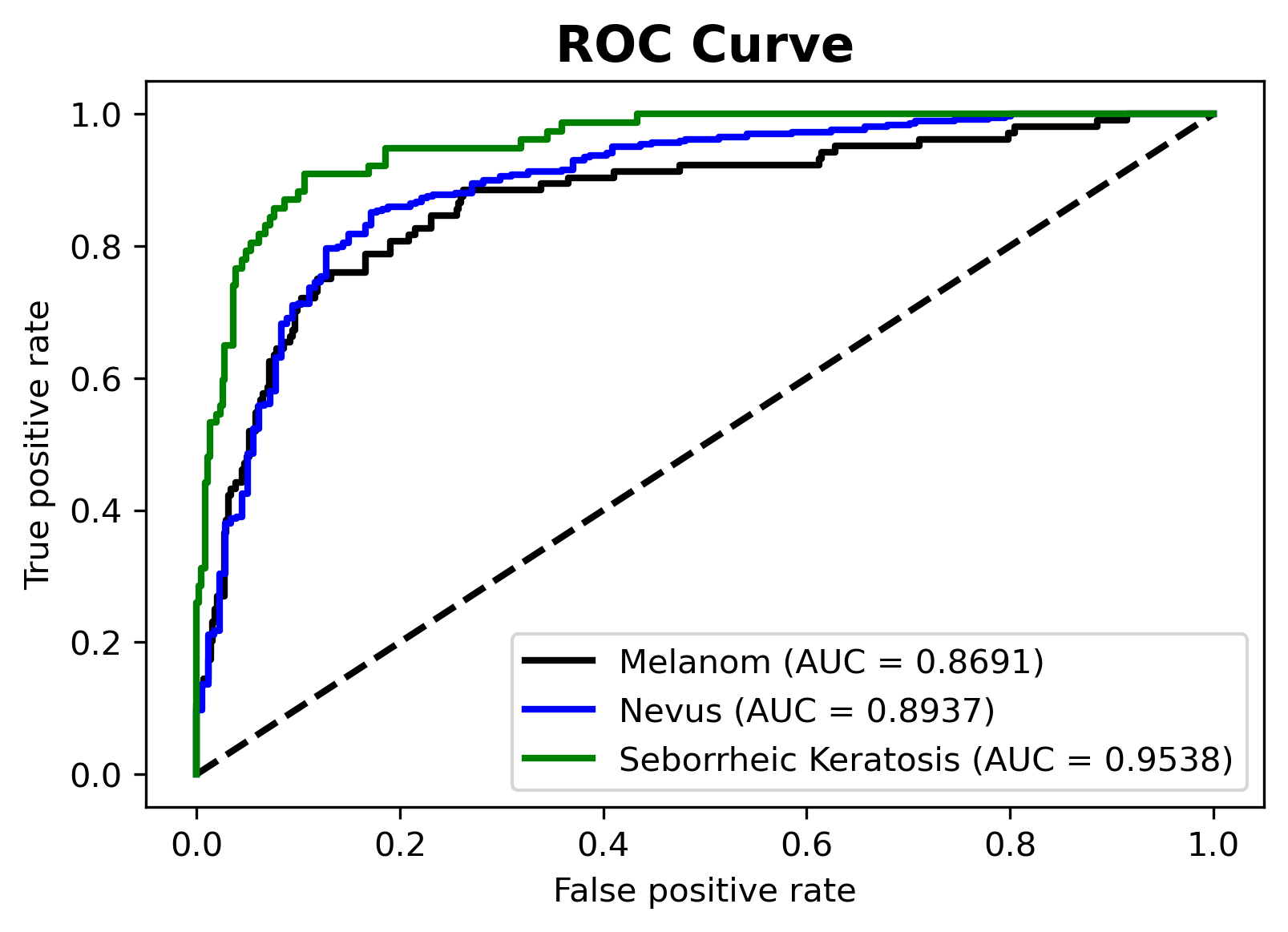}
\caption{ROC Curve for MobileNet with 16 Batch Size}
\end{figure}

\begin{figure}[h!]
\centering
\label{fig:mobilenetv1_32}
\includegraphics[width=8cm]{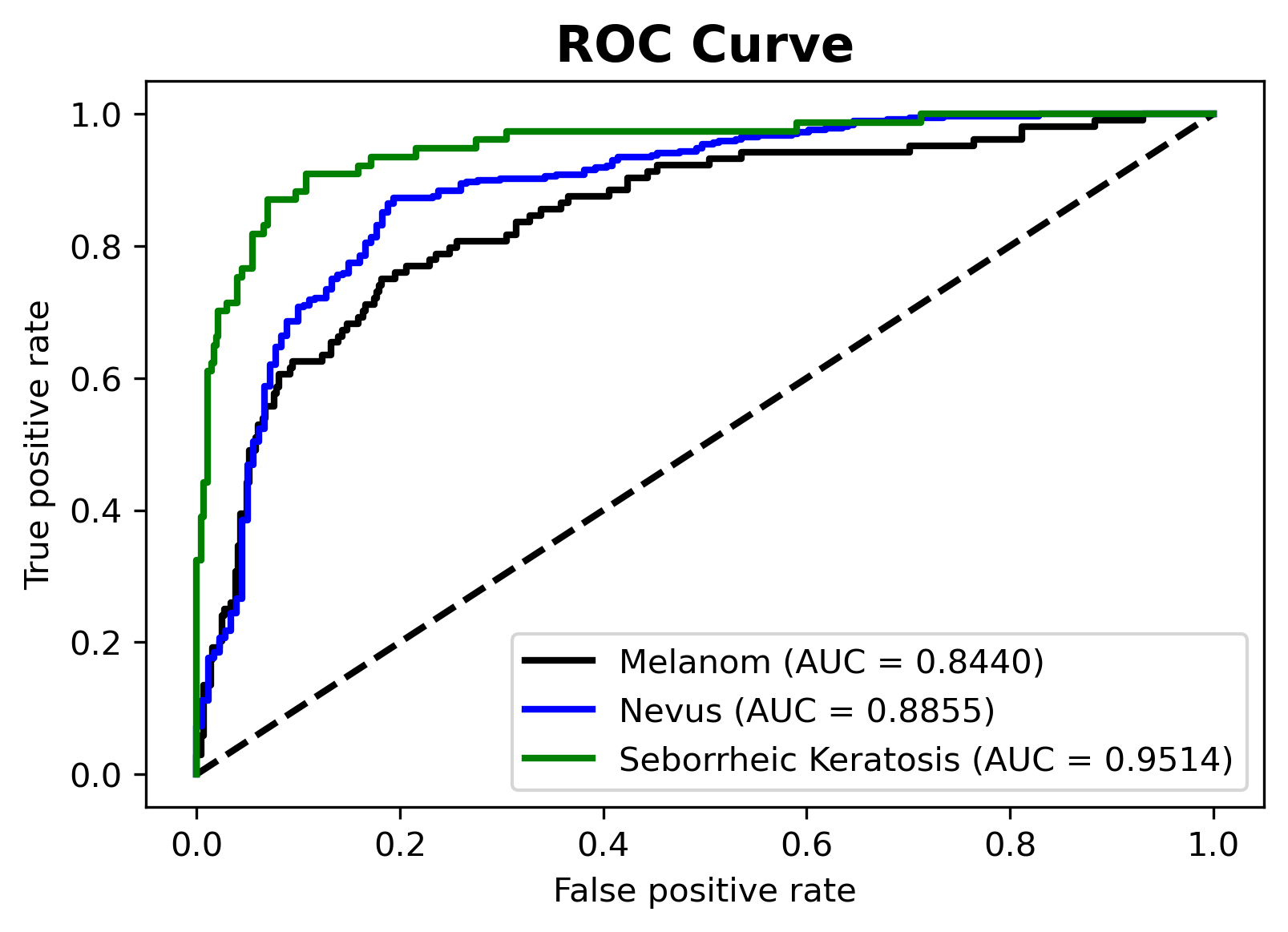}
\caption{ROC Curve for MobileNet with 32 Batch Size}
\end{figure}

\begin{figure}[h!]
\centering
\label{fig:mobilenetv1_64}
\includegraphics[width=8cm]{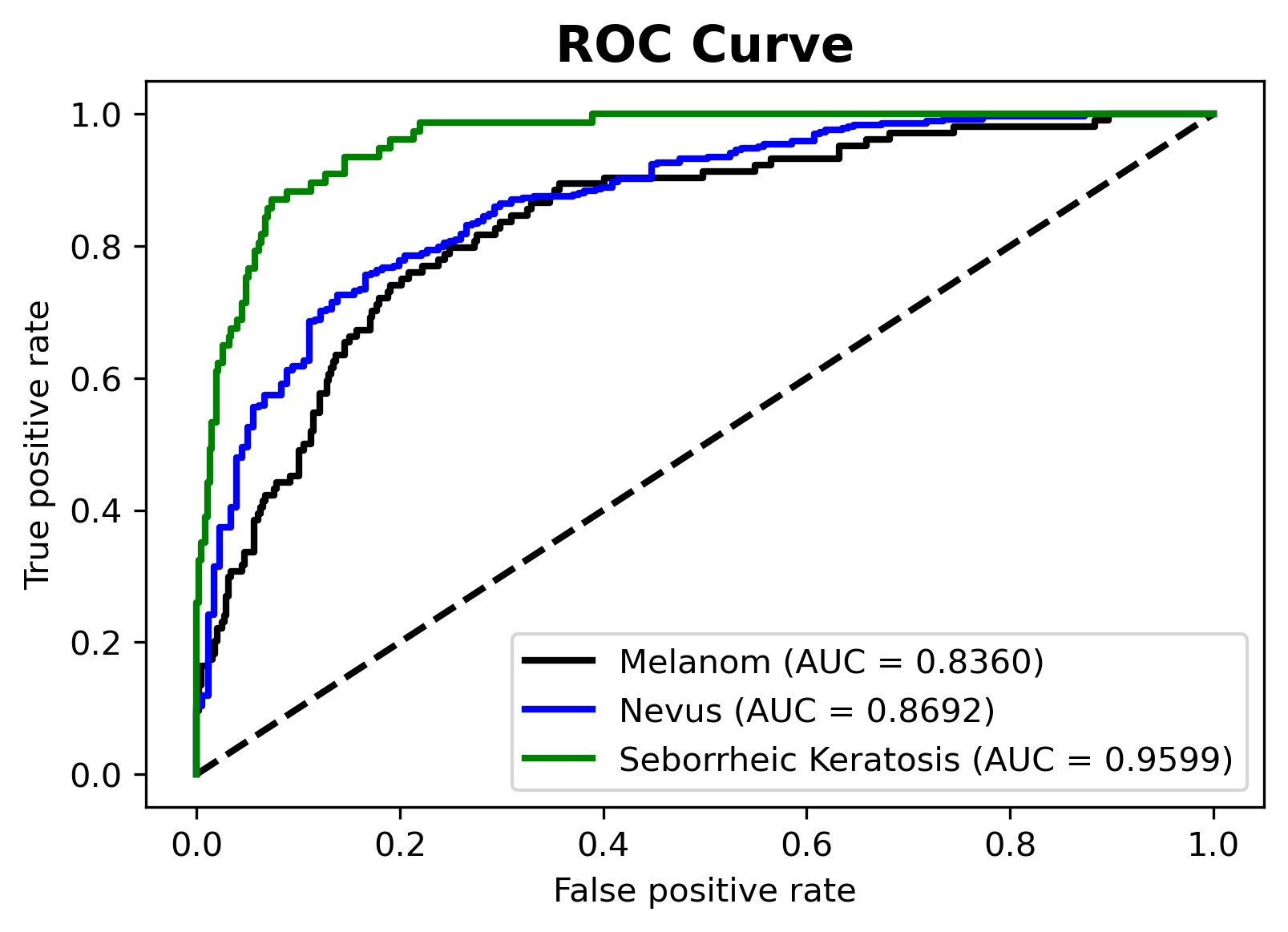}
\caption{ROC Curve for MobileNet with 64 Batch Size}
\end{figure}

\begin{figure}[h!]
\centering
\label{fig:mobilenetv2_16}
\includegraphics[width=8cm]{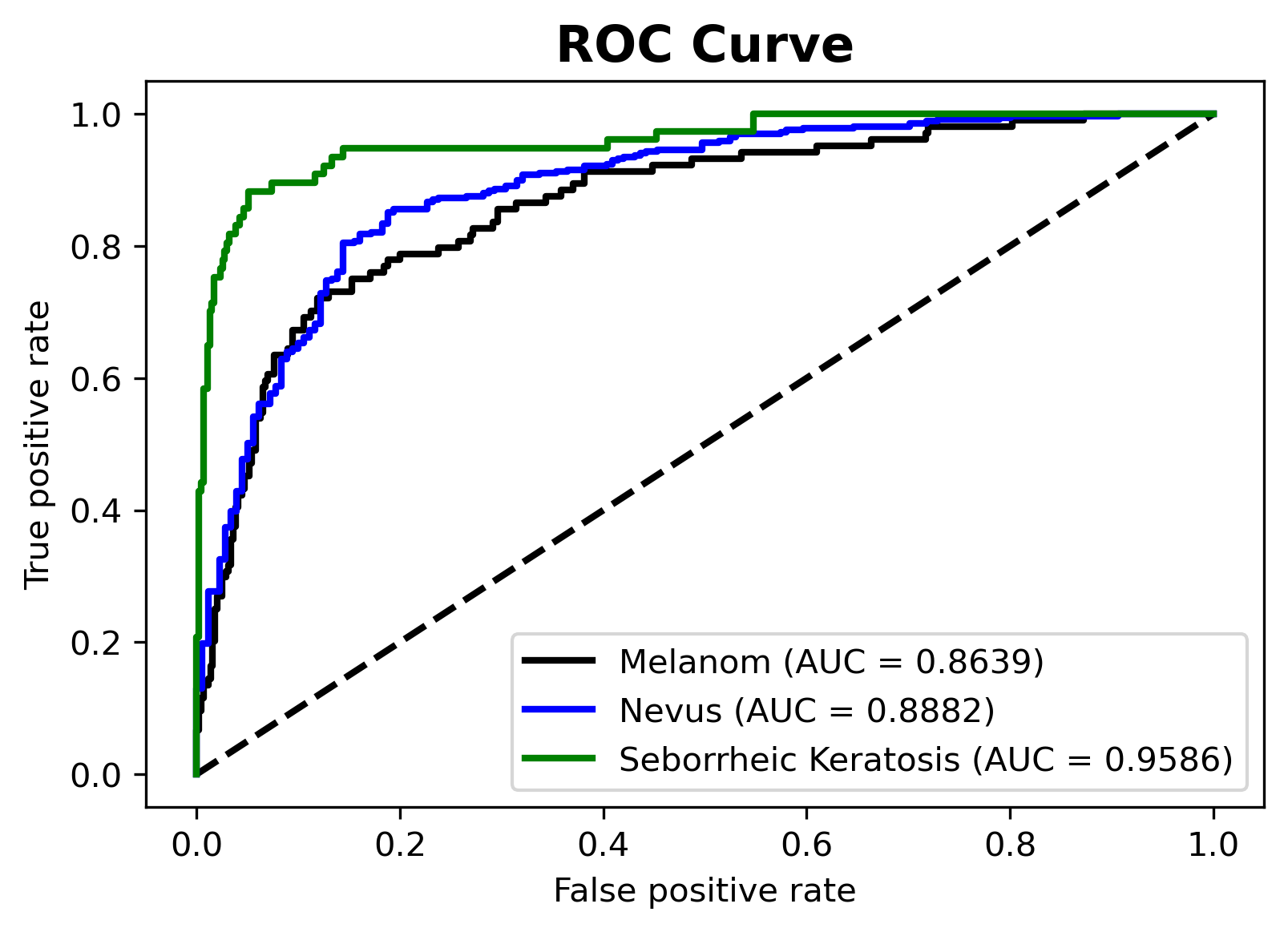}
\caption{ROC Curve for MobileNetV2 with 16 Batch Size}
\end{figure}

\begin{figure}[h!]
\centering
\label{fig:mobilenetv2_32}
\includegraphics[width=8cm]{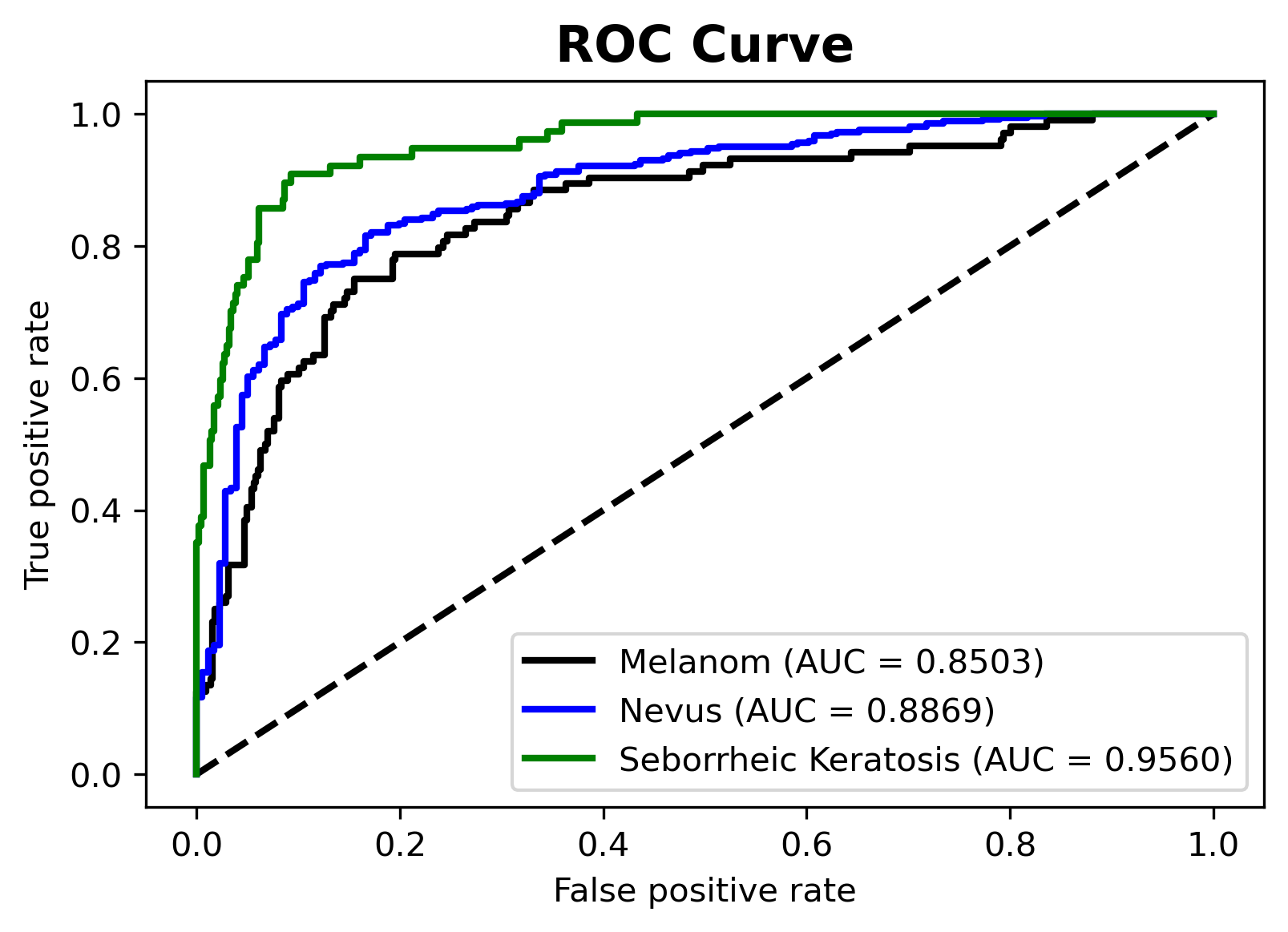}
\caption{ROC Curve for MobileNetV2 with 32 Batch Size}
\end{figure}

\begin{figure}[h!]
\centering
\label{fig:mobilenetv2_64}
\includegraphics[width=8cm]{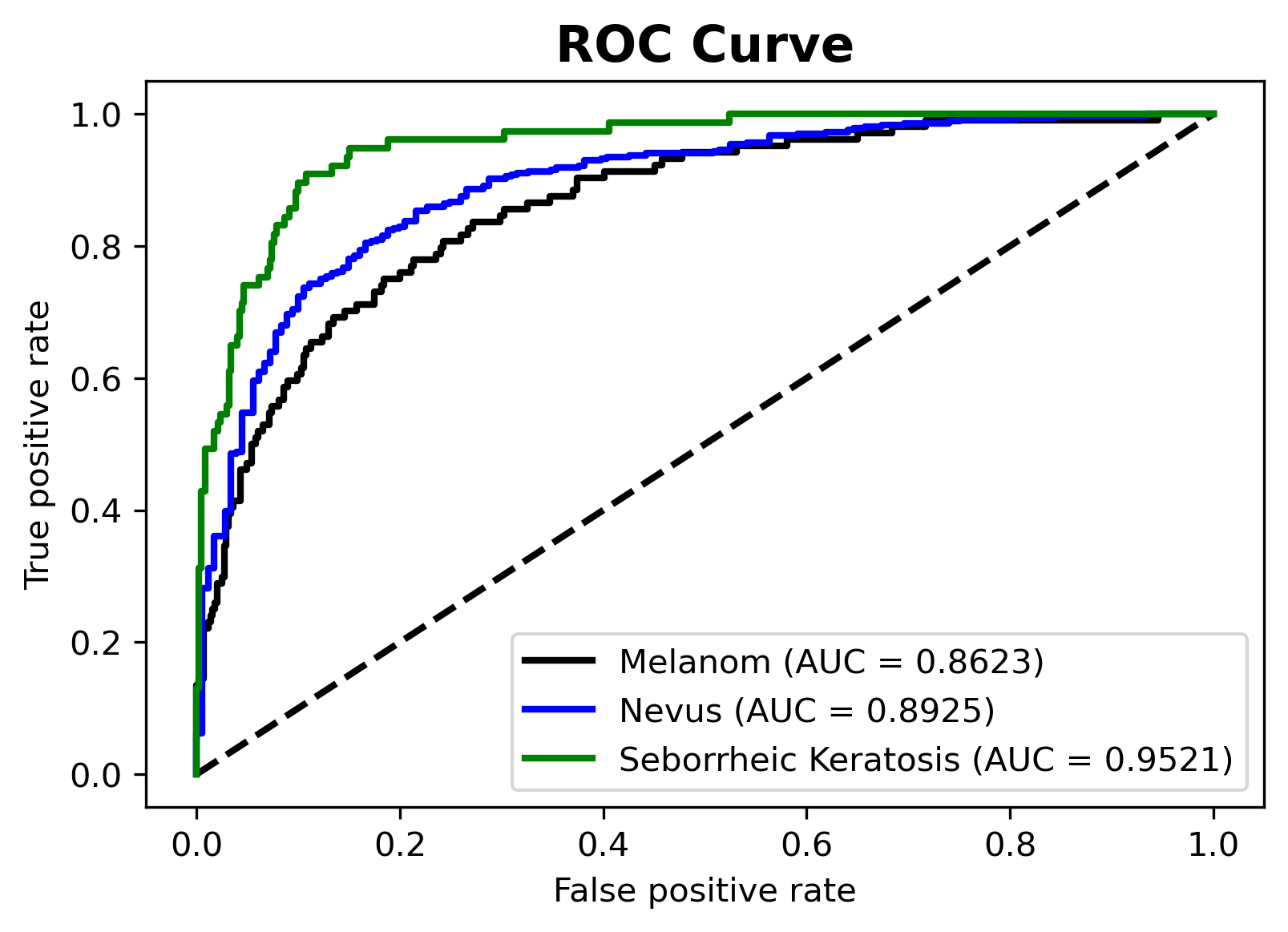}
\caption{ROC Curve for MobileNetV2 with 64 Batch Size}
\end{figure}

\begin{figure}[h!]
\centering
\label{fig:nasnet_16}
\includegraphics[width=8cm]{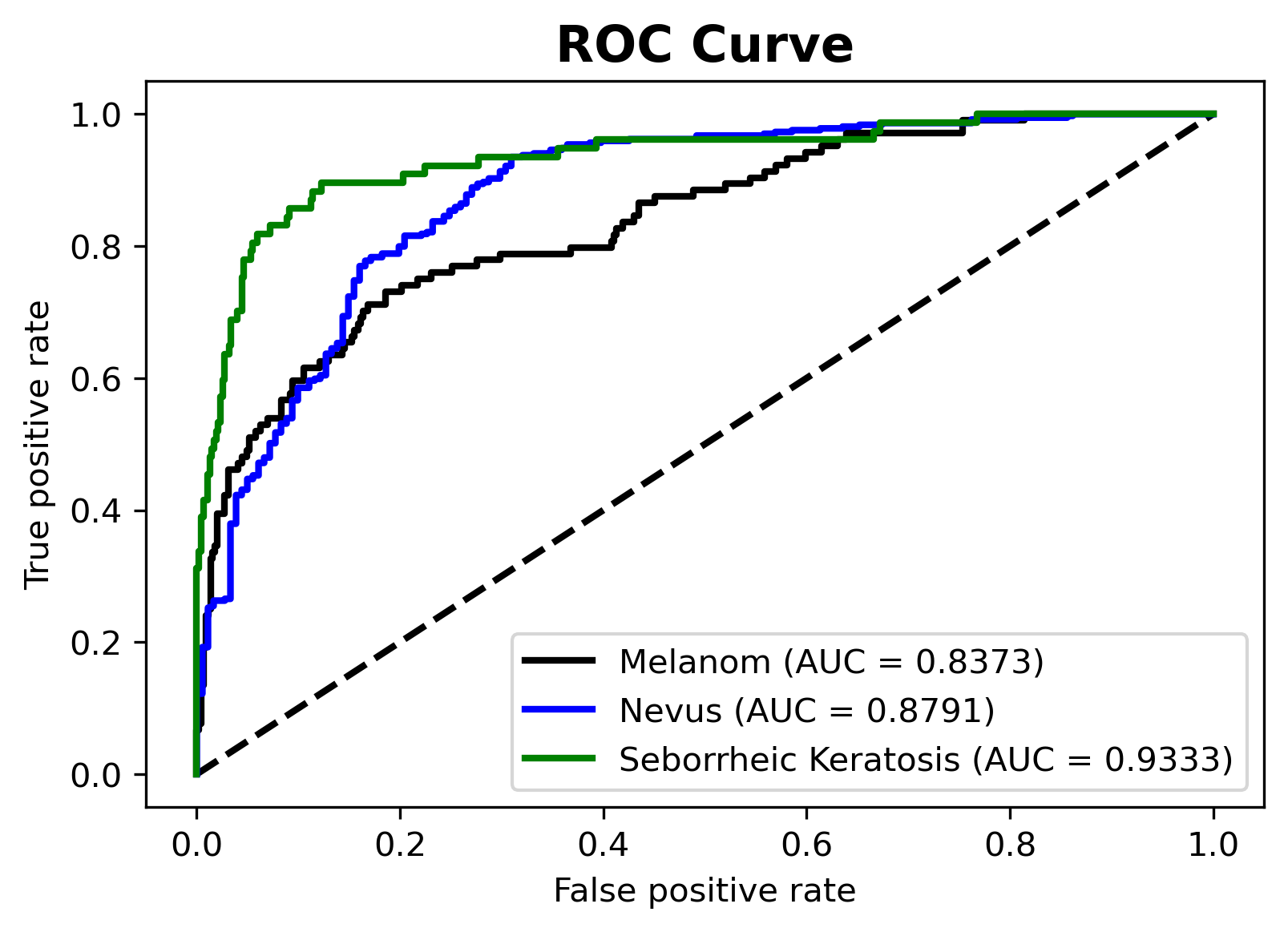}
\caption{ROC Curve for NASNetMobile with 16 Batch Size}
\end{figure}

\begin{figure}[h!]
\centering
\label{fig:nasnet_32}
\includegraphics[width=8cm]{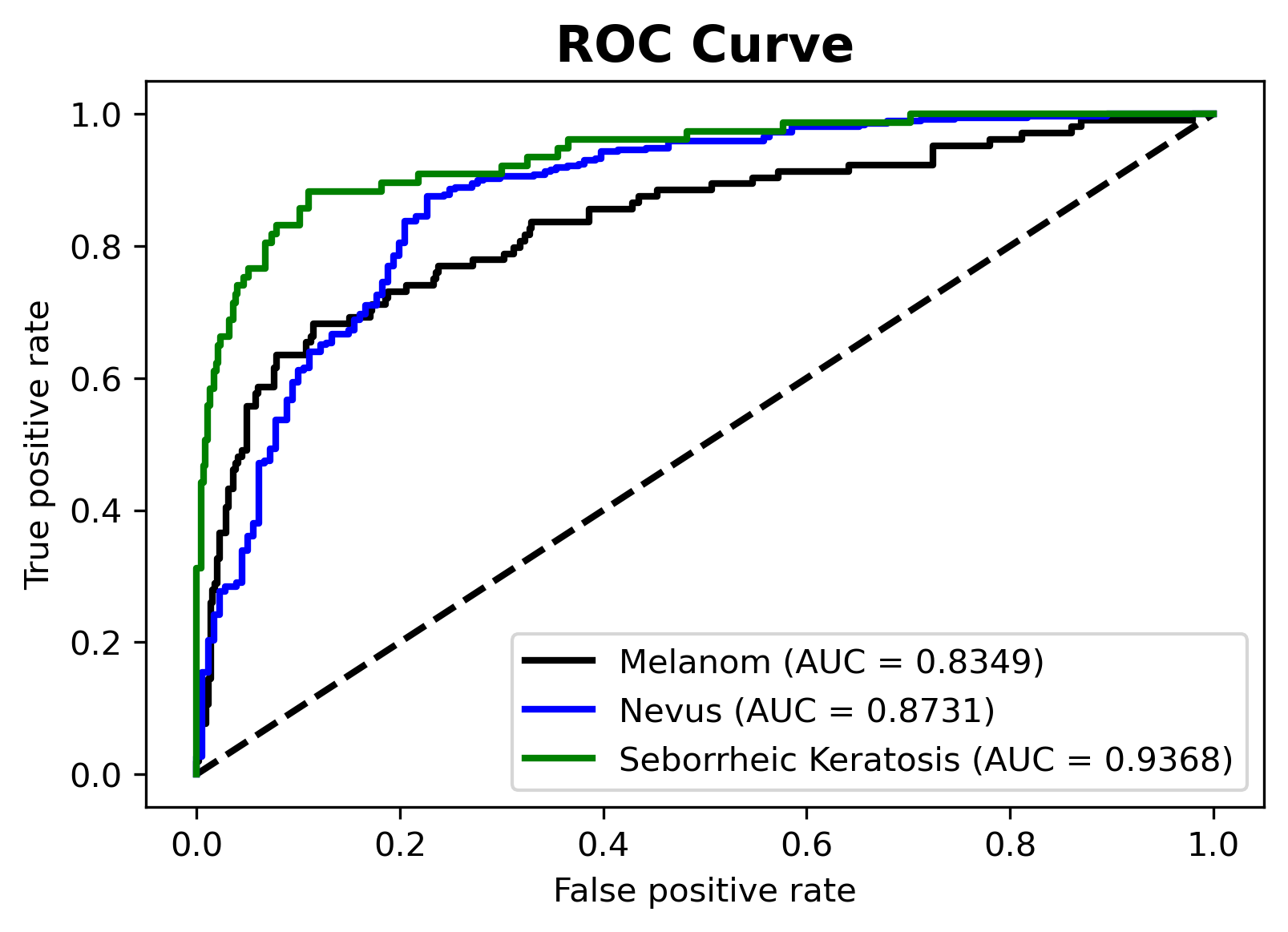}
\caption{ROC Curve for NASNetMobile with 32 Batch Size}
\end{figure}

\begin{figure}[h!]
\centering
\label{fig:nasnet_64}
\includegraphics[width=8cm]{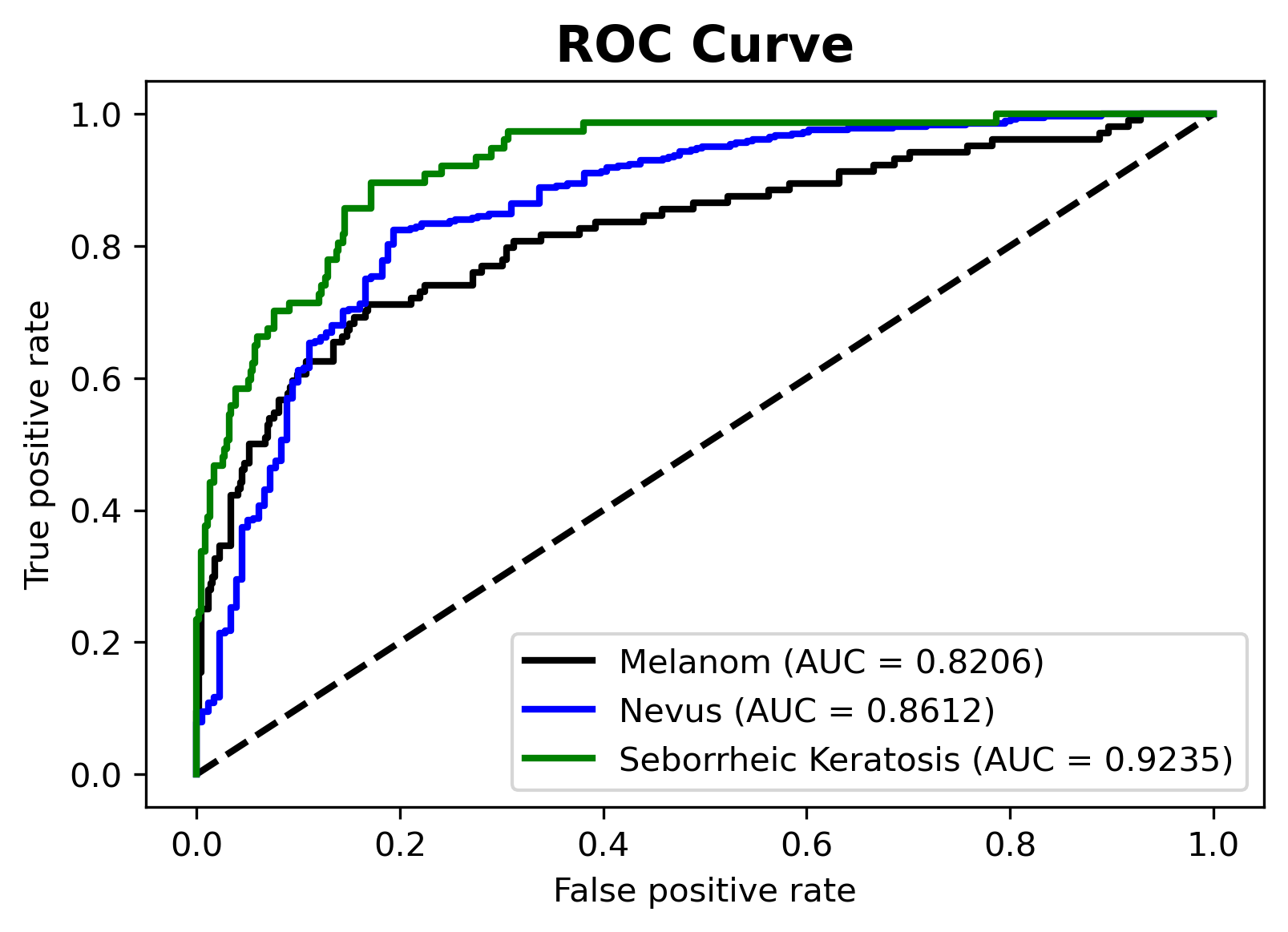}
\caption{ROC Curve for NASNetMobile with 16 Batch Size}
\end{figure}

\end{document}